# Observation of giant magnetocaloric effect in HoCoSi

Sachin Gupta and K.G. Suresh[*]

*Department of Physics, Indian Institute of Technology Bombay, Mumbai-400076, India*

## Abstract

We report the magnetic and magnetocaloric properties of HoCoSi, which shows second order ferro-para magnetic transition at 14 K. It is found to exhibit a large magnetic entropy change of 13 J/kg K for a field change of 20 kOe, which increases to 20.5 J/kg K at 50 kOe. The refrigerant capacity is found to be 135 and 410 J/kg for 20 and 50 kOe, respectively. The large, low-field adiabatic temperature change (~ 3 K) and a large refrigerant capacity make this compound a suitable candidate for magnetic refrigeration applications at around 15 K.

**Keywords:** Magnetic materials; Metals and alloys; Functional; Magnetocaloric effect

[*]Corresponding author (email: suresh@phy.iitb.ac.in)

Tel. +91-22-25767559, Fax:   +91-22-25723480

# 1. Introduction

In the last two decades, there has been an intensive search for potential materials that can be used as magnetic refrigerants. The thrust on this topic stems from the disadvantages such as low efficiency, environmental problems, large power consumption, adaptability etc. associated with the conventional gas compression/expansion method [1]. Researchers have paid much attention in this field and explored a range of potential magnetic materials with large magnetocaloric effect (MCE) and refrigerant capacity (RC). The magnetic refrigeration potential is based on MCE and can be measured in terms of isothermal magnetic entropy change ($-\Delta S_M$) and/or adiabatic temperature change ($\Delta T_{ad}$). The study of MCE not only helps in identifying potential magnetic refrigerant materials, but also leads to a better understanding of the fundamental physical properties of magnetic materials [2,3]. It has been observed that the magnetic materials with first order magnetic phase transition, such as $Gd_5(Si,Ge)_4$ [4], $Ni_2MnGa$ [5], MnAs [6] show large MCE around their magnetic ordering temperatures. But large hysteresis losses existing in materials with first order phase transition make magnetic refrigerants less efficient in a refrigeration cycle. Because of this, the search for new potential materials with second order transition and showing negligible thermal and field hysteresis is on for some time. Discovery of such materials will offer a big leap in the realization of magnetic refrigeration technology.

RTX (R=rare earth, T=transition metal, and X= p block element) family comprises of a number of compounds having different crystal structures showing diverse and interesting magnetic and electrical properties. Some compounds of this family show large MCE [7,8]. Motivated by these results, we have studied HoCoSi compound belonging to this family and the results are presented in this paper. Here we focus on the magnetic and magnetocaloric properties of polycrystalline HoCoSi.

## 2. Experimental Details

Polycrystalline HoCoSi compound was synthesized by arc melting stoichiometric proportion of Ho (99.9% purity), Co and Si (99.99% purity) in water cold copper hearth under argon atmosphere. The ingot was remelted several times for better mixing of constituent elements. As-cast sample was sealed in evacuated quartz tube and annealed for 100 h at 800 ℃ to improve the homogeneity. The phase purity of annealed sample was checked by X'PERT PRO diffractometer at room temperature using Cu Kα (λ= 1.54 Å) radiation. Magnetic measurements, M(T) and M(H), were performed on Quantum Design physical property measurement system (PPMS 6500). The heat capacity C(T) measurement was carried out in a PPMS using thermal relaxation method.

## 3. Results and discussion

The Rietveld refinement of the XRD pattern taken at room temperature shows that the compound is single phase and crystallizes in TiNiSi type orthorhombic crystal structure with space group Pnma (No. 62). The lattice parameters obtained from the refinement are a= 6.80 Å, b= 4.13 Å, c= 7.11 Å and are close to the reported values [9].

The temperature dependence of dc susceptibility (χ) data taken in zero-field-cooled (ZFC) and field-cooled (FC) modes under a field of 500 Oe is shown in the inset (a) of Fig. 1. The compound shows reversible behavior in the transition region, but shows small thermo-magnetic irreversibility at very low temperatures, which may arise due to the domain wall pinning effect. The inverse susceptibility is shown on the right-hand panel of inset (a) of Fig. 1. and fitted with Curie-Weiss law in the paramagnetic region. The effective magnetic moment ($\mu_{eff}$) and the paramagnetic Curie temperature ($\theta_p$) estimated from the fit are found to be 10.9 $\mu_B$/f.u. and 6 K respectively. The value of $\mu_{eff}$ is close to that of free $Ho^{3+}$ ion moment (= 10.6 $\mu_B$), suggesting that 4f electrons are well localized and that there is no appreciable moment from Co. The inset (c) of Fig. 1 shows the temperature derivative of χ, giving a negative peak, indicating ferromagnetic ordering at 14 K (termed as $T_C$), which is close to that reported in Ref [10].

The magnetization isotherms taken at different temperatures in an interval of 2 K and upto 50 kOe field are shown in Fig. 1. The magnetization is not saturated at 50 kOe but it shows a tendency to saturate at higher field. The significant change in the slope of magnetization isotherms near $T_C$ is expected to result in significant magnetic entropy change. The inset (b) of Fig. 1 shows the magnetization isotherm at 2 K in the increasing and the decreasing field modes. It is clear from the inset that the compound shows no magnetic hysteresis, which indicates perfect magnetic reversibility or soft ferromagnetic nature, which is very advantageous for magnetic refrigeration applications.

The order of magnetic phase transition is usually revealed by the Arrott plots. Therefore we have plotted the Arrott plots at different temperatures. According to the Banerjee criterion [11], the plot of $M^2$ vs. H/M shows negative slope for first order magnetic phase transition and it is positive for second order phase transition. In Fig. 2 one can see that the plots show very small negative slopes at the initial fields and low temperatures, which gives a hint of first order magnetic phase transition. But the absence of significant hysteresis in the M-H data and the λ-shape peak in heat capacity (not shown here) rule out first order transition. Therefore, the nature of magnetic transition was further analyzed using the Inoue-Shimizu model [12], which involves a Landau expansion of magnetic free energy, F(M,T) up to sixth power of magnetization (M) and can be expressed as

$$F(M,T) = \frac{a_1(T)}{2}M^2 + \frac{a_3(T)}{4}M^4 + \frac{a_5(T)}{6}M^6 + .... - \mu_0 HM. \qquad (1)$$

The parameters $a_1(T)$, $a_3(T)$, and $a_5(T)$ are the Landau coefficients and can be estimated from the equation of state given by

$$\mu_0 H = a_1(T)M + a_3(T)M^3 + a_5(T)M^5. \qquad (2)$$

The temperature dependence of $a_1(T)$ and $a_3(T)$ gives two characteristic temperatures $T_C$ and $T_0$ respectively, which help us to identify the nature of a magnetic transition. If the sign of $a_3(T_C)$ is negative, it refers to first order transition, while a positive value indicates second order transition. The insets (b and c) in Fig. 3 show the temperature dependence of $a_1(T)$ and $a_3(T)$. The minimum of $a_1(T)$ is marked as $T_C$ and found to be 12 K. $T_0$ is the temperature at which $a_3(T)$ passes from zero and is found to be around $T_C$. Thus, one can see that in the present case the value of $a_3(T)$ is positive at $T_C$, which confirms that the magnetic transition is second order type [13].

The $\Delta S_M$ has been calculated from the magnetization (M-H-T) data using Maxwell's relation, $\Delta S_M = \int_0^H (\partial M / \partial T)_H dH$. One can see from inset of Fig. 2 that the height and the width of -ΔS$_M$ vs. T curve increase with field. The peak value of -ΔS$_M$ becomes 13 and 20.5 J/kg K for fields of 20 and 50 kOe respectively.

The estimation of ΔS$_M$ and ΔT$_{ad}$ from the heat capacity data (C-H-T) has been carried out using following the thermodynamic equations [14]

$$\Delta S_M(T,H) = \int_0^T \frac{C(T',H) - C(T',0)}{T'} dT' \qquad (3)$$

$$\Delta T_{ad}(T)_{\Delta H} \cong [T(S)_{H_f} - T(S)_{H_i}]_S \qquad (4)$$

The main panel in Fig. 3 depicts the MCE plot calculated from the C-H-T data at 10 and 20 kOe fields. One can note that there is a difference in the peak values of -ΔS$_M$ (at 10 and 20 kOe) calculated from the M-H-T and the C-H-T data. It has been observed that the peak value of -ΔS$_M$ for HoCoSi is comparable to many potential refrigerant materials as shown in Table I. The inset (a) of Fig. 3 shows ΔT$_{ad}$ variation, which shows that the value at 20 kOe is 3.1 K.

The RC is defined as the amount of heat transferred between hot and cold reservoirs during an ideal refrigeration cycle. The values of RC calculated as the product of $\Delta S_M^{max}$ and full width at half maximum in ΔS$_M$(T) curve (RC=$-\Delta S_M^{max} \delta T_{FWHM}$) [2] is found to be 135 and 410 J/kg for 20 and 50 kOe. From the comparison of the MCE parameters (-ΔS$_M$, RC and ΔT$_{ad}$) in Table I, one can see that HoCoSi shows promising low-field MCE.

## 4. Conclusions

In conclusion, HoCoSi shows second order ferromagnetic to paramagnetic transition. The compound shows giant reversible MCE with a maximum magnetic entropy change of 20.5 J/kg K at 14 K for a field of 50 kOe. The RC is comparable to that of many potential refrigerant materials. Another important advantage is that the compound is magnetically soft, which minimizes the energy loss when cycling the field in a magnetic refrigerator. Relatively large and low-field -$\Delta S_M^{max}$, ΔT$_{ad}$, and RC values, along with magnetic softness makes this compound an attractive magnetic refrigerant suitable in the low temperature regime.

# Acknowledgment

SG is thankful to CSIR, Govt. of India for providing research fellowship.

**Figure Captions:**

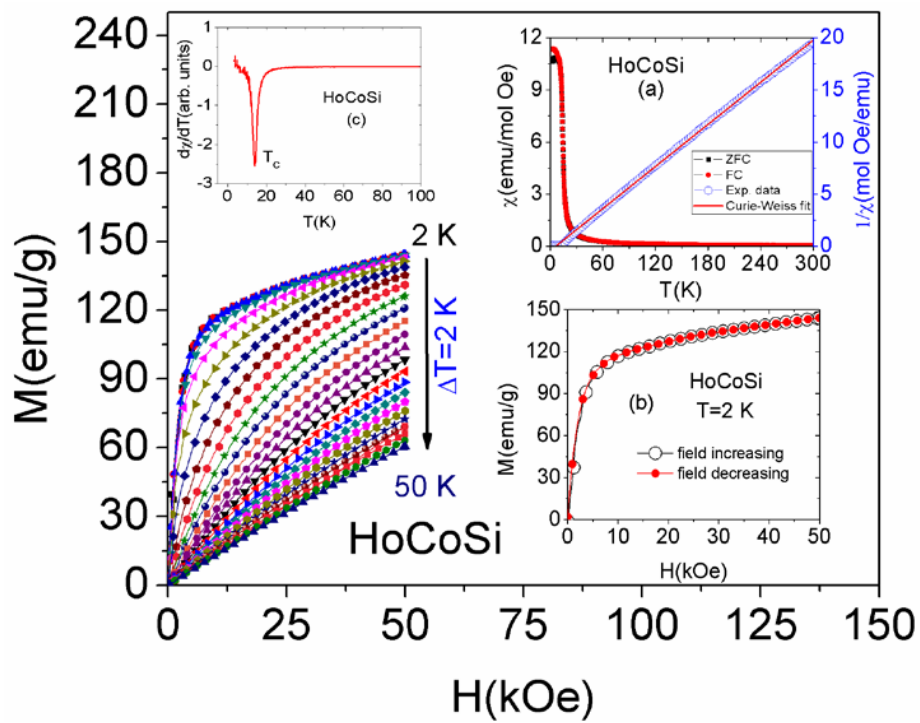

**Fig. 1.** Magnetization isotherms taken at different temperatures for HoCoSi. Insets : (a) $\chi$ vs. T plot (left-hand scale) and Curie-Weiss fit to inverse susceptibility (right-hand scale), (b) M(H) curve at 2 K with increasing and decreasing field modes, (c) derivative of $\chi(T)$.

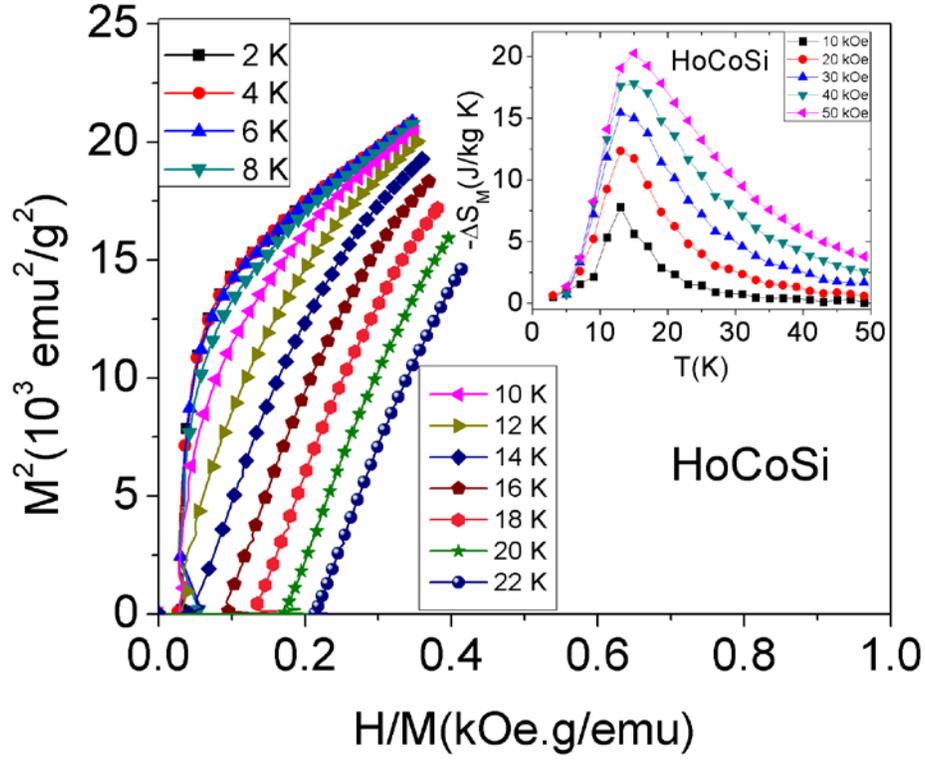

**Fig. 2.** Arrott plots at some selected temperatures. Inset shows -$\Delta S_M$ vs. T plots at different fields.

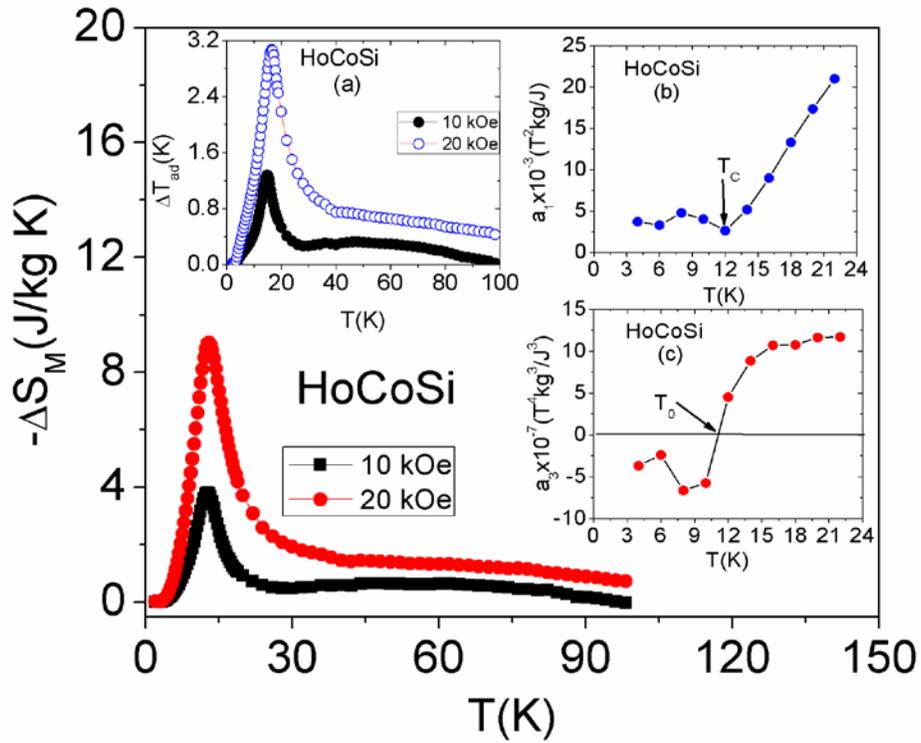

**Fig. 3.** -$\Delta S_M$ vs. T at 10 and 20 kOe field calculated from C-H-T data. Insets: (a) $\Delta T_{ad}$ vs. T plots at 10 and 20 kOe field, temperature dependence of Landau coefficients (b) $a_1(T)$ and (c) $a_3(T)$.

**Table I.** The values of $T_{C,N}$, -$\Delta S_M$, RC and $\Delta T_{ad}$ for some well known magnetic refrigerant materials including HoCoSi (this work).

| Compound | $T_C$ or $T_N$ | -$\Delta S_M$ (J/kg K) at 50 kOe | RC (J/kg) at 50 kOe | $\Delta T_{ad}$ (K) at 20 kOe | Reference |
|---|---|---|---|---|---|
| HoCoSi | 14 | 20.5 | 410 | 3.1 | This work |
| ErFeSi | 22 | 23.1 | 365 | 2.9 | 7 |
| ErRuSi | 8 | 21.2 | 416 | - | 8 |
| Er$_3$Ni$_2$ | 17 | 19.5 | 407 | 3.3 | 15 |
| ErMn$_2$Si$_2$ | 4.5 | 25.2 | 365 | 5.4 | 16 |
| PrCo$_2$B$_2$ | 16 | 8.1 | 104 | 4.3 | 2 |